\documentclass[aps,prl,floats,twocolumn,showpacs]{revtex4}
\usepackage[dvips]{graphicx,color}
\begin{document}
\title{Shot noise of a quantum dot with non-Fermi liquid correlations}
\author{Alessandro Braggio$^{1}$, Rosario Fazio$^{2}$ and Maura Sassetti$^{1}$
\vspace{1mm}}
\affiliation{$^{1}$Dipartimento di Fisica, INFM-Lamia, 
  Universit\`{a} di Genova, Via Dodecaneso 33, I-16146 Genova, 
  Italy\\
  $^{2}$ NEST-INFM $\&$ Scuola Normale Superiore, I-56126 Pisa, Italy 
\vspace{3mm}} 
\date{\today}

\begin{abstract}
    The shot noise of a one-dimensional wire 
  interrupted by two barriers shows interesting features related to
  the interplay between Coulomb blockade effects, Luttinger
  correlations and discrete excitations. At small bias 
  the Fano factor reaches the lowest attainable value, $1/2$, 
  irrespective of the ratio of the two junction resistances.
  At larger voltages this asymmetry is power-law renormalized
  by the interaction strength. We discuss how the 
  measurement of current and these features of the noise
  allow to extract the Luttinger liquid parameter. 
\end{abstract}

\maketitle 
A comprehensive understanding of the transport properties of mesoscopic 
conductors can be achieved with the study of both the average current and 
its fluctuations. 
Unlike equilibrium noise, linked to the conductance by the
fluctuation-dissipation theorem, shot noise provides further understanding 
not related to the average current. Noise is typically very sensitive to 
correlation mechanisms arising either due to statistics, interference 
phenomena and interactions~\cite{deJong97,blanter00}. 
One of the most striking effects 
recently observed is the measurement~\cite{depicciotto97,saminadayar97} 
of the fractional charge in the Fractional Quantum Hall Effect.

The understanding of the role of electron interaction on shot noise is
far from being complete. At the present most of the attention is
devoted to two classes of systems, mesoscopic junctions in the Coulomb
blockade regime and one dimensional (1D) wires where
electron interaction lead to a Luttinger liquid behaviour.
In single electron tunnelling transistors noise has been studied
both in the sequential
tunnelling~\cite{davies92,hershfield93,hanke93,korotkov94} and in the
cotunnelling~\cite{averin00,sukhorukov00} regime. Deviation of the
shot noise from the Poisson value was predicted when two charge
states are almost degenerate. As a function of the transport voltage,
the Fano factor shows dips in correspondence of the steps of the
Coulomb staircase.  Experiments on the shot noise suppression due to
Coulomb blockade effects has been performed by Birk 
{\em et al.}~\cite{birk95} with a good agreement with the theoretical
prediction. As a function of the gate voltage, shot noise exhibits a
periodic suppression below the Poisson value~\cite{hanke93}.

Another situation where interactions have a drastic
consequence~\cite{v95} is in 1D systems. Here, there are no
fermionic quasiparticle, and the low energy excitations
consist of independent long-wavelength oscillations of the
charge and spin density, which propagate with different velocities.
The density of states has a power-law behaviour and the 
transport properties cannot be described in terms of the conventional
Fermi-liquid approach.  A Luttinger liquid (LL) with an arbitrarily
small barrier leads to a complete suppression of transport at low
energies~\cite{kane92}.  By now there are several works which show the
emergence of a LL behaviour. These studies embrace a
variety of physical systems as quantum wires~\cite{auslaender00},
carbon nanotubes~\cite{bockrath99} and edge state in the Fractional
Quantum Hall effect~\cite{milliken96}. Shot noise in systems with LL
correlations provides important informations and a number of very
interesting properties have been found so far.  
In addition  to the possibility to measure the
fractional charge, interactions renormalize the singularity 
at the ``Josephson'' frequency~\cite{chamon96}.  Other
clear signatures include the complete locking of
the shot noise in coupled LL~\cite{trauzettel02}.
We thus expect that 
a single electron transistor made of a 1D system will have 
a rather rich phenomenology.

In this Letter we study the shot noise of a
quantum dot, formed by two barriers, in a 1D wire with LL
correlations in the sequential tunnelling regime.  
We consider the quantum dot with discrete energy excitations,
ranging from the ``single resonant level'' to a continuum distributions.
Beside the suppression of the shot noise due to the Coulomb 
blockade~\cite{hershfield93}, we find additional effects induced 
by the non-Fermi liquid correlations. These last features are what
we are mainly looking at.  We show that the shot noise reveals several
unambiguous evidences of Luttinger correlations  and presents 
new and complementary informations with respect to the current
measurements. The results  can be used to test 
independently the non-Fermi liquid nature of the system.

The Hamiltonian of the system is composed by the
term which describes electrons in the quantum wire and the scattering
induced by the presence of the two barriers.  In the bosonized form
the part due to an homogeneous spinless interacting one-dimensional
electron gas is expressed in terms of an harmonic field operators
($\hbar =1$) $[\Pi(x),\Theta(x')]=i\delta(x-x')$\cite{v95}
\begin{equation}
\label{hamiltonian}
H_{0}=\frac{v_{\rm F}}{2}\int {\rm d}x\,\left\{\Pi^{2}(x)
+\frac{1}{g^{2}}[\partial_{x}\Theta(x)]^{2}\right\}\,.
\end{equation}
The interaction parameter is $g^{-2}=1+V(q= 0)/\pi
  v_{\rm F}$ with $v_{\rm F}$ the Fermi velocity and $V(q)$ the
Fourier transform of the electron-electron interaction.  The excitations of
Eq.(\ref{hamiltonian}), at fixed average density
$\rho_0=k_{\rm F}/\pi$, are collective density waves with
dispersion $\omega(q)=v_{\rm F}|q|/g$.
The scattering is due to the left (l) and to the right (r) 
barriers at positions $x=\pm d/2$ with different heights 
$U_{\rm l,r}$. 
The corresponding tunnelling Hamiltonian is~\cite{kane92}
\begin{equation}
\label{barrier}
H_{\rm T}=\sum_{p=\pm}\!(U_{\rm l}+pU_{\rm r})
\prod_{s=\pm}\cos{\pi(N^{s}\!+\!n_{0}\delta_{s,-}-\frac{1-p}{4})}
\end{equation}
where 
$N^{\pm}=[\Theta(d/2) \pm\Theta(-d/2)]/\sqrt{\pi}$, with $N^{-}$
the deviation of the number of electrons in the dot from the mean
value, $n_0=d\rho_{0}$, and $N^{+}/2$ the imbalanced electrons between
the left and right leads.  The coupling to the source-drain bias $V$
and the gate voltage $V_{\rm g}$ is given by $H_{\rm V}= -e(VN^{+}/2+
\alpha V_{\rm g}N^{-})$, with $\alpha$ the ratio between gate and total 
capacitances.

The zero-frequency power spectrum of the noise is given by
the current-current correlation function
\begin{equation}
\label{noise}
S=\int\limits_{-\infty}^{\infty}dt
\left[\left\langle\delta I(t)\delta I(0)\right\rangle
+\left\langle\delta I(0)\delta I(t)\right\rangle
\right]\,,
\end{equation}
where $\delta I(t)\equiv I(t)-\langle I\rangle$ is the (total) current
fluctuation operator with respect to the average $\langle I\rangle$.
Since we are interested in the zero-frequency noise, due to continuity
equation and simmetries of the correlation functions, it is sufficient
to study fluctuations in the current through the right 
or the left junctions $I_{\rm
  r/l}(t)=e\dot{\Theta}(\pm\frac{d}{2},t)/{\sqrt\pi}$.

The current fluctuations are determine by the
dynamics of the variables $N^{\pm}$ under the influence of the
electrostatic fields in the tunnelling potential (\ref{barrier}).  We
consider the sequential tunnelling regime, neglecting contributions
from coherent processes. This assumption is valid in the weak
tunneling regime for not too low temperatures~\cite{braggio01}. 
This is exactly the regime relevant for  available experiments.  For
high barriers the dynamics is dominated by tunnelling through the left
and right junctions with resistances $R_{\rm l}$ and $R_{\rm r}$
respectively.  There are two characteristic energy scales. 
The energy $E_{\rm c}$ has to be supplied in order to
transfer one extra charge to the island from the neutral
configuration. It determines the dot ground state chemical potential 
with $n$ electrons $\mu(n)=2E_{\rm c}(n-n_{\rm g})$, with
$n_{\rm g}=n_0+\alpha eV_{\rm g}/2E_{\rm c}-1/2$. A
discrete energy $\varepsilon$ is necessary to further excite the
discrete plasmon modes inside the dot.  An explicit expression of
these two energy scales depends on the microscopic and circuital
details of the system~\cite{braggio01}.  The coupling to gates and the
presence of possible long range electron interactions tend to modify
these values~\cite{k00}. For these reasons we will treat them as
independent parameters.

The shot noise is obtained from the solution of a time dependent
master equation for the occupation probabilities of charges in the dot
as described in Ref.~\onlinecite{hershfield93}. The new ingredients
here are the tunnelling rates that depend on the Luttinger
correlations~\cite{braggio01}. 
For a tunnelling process, to which is associated a change $E$ in the
electrostatic energy, the rate is proportional to
$\gamma(E)=\sum_{p=0}^{\infty} \Gamma(p+1/g)f(E-p\varepsilon,g)/p!  $,
with $f(E,g)={\rm e}^{E/2T} |\Gamma\left(1/(2g)+iE/2\pi T\right)|^2$
($k_{\rm B}=1$).
Note that this expression is obtained by assuming a thermal distribution
for the plasmons in the leads and in the dot with
$T\ll\varepsilon$.  This implies that the initial dot state is always
the ground state. The plasmon excitations are present only in the
final states. Such a description is appropriated when the relaxation
processes are faster than the tunnelling processes. In this case any
excited state will decay into the ground state long before the next
electron tunnels in the dot.  The possibility to tune
the external bias voltage in the energy $E$ allow to investigate the
discreteness of the plasmon energy, ranging from the
``single resonant level'' ($E<\varepsilon$) where no plasmon are
excited, to the continuum of plasmon distribution
($E\gg\varepsilon$).  In these two extreme cases the rates acquire
the simplifed form $\gamma(E)\approx f(E,g)$ and
$\gamma(E)\approx f(E,g/2)$ respectively. In the
first case the power law correlations are only due to 
the Luttinger liquid nature of the leads. In the latter the dot
itself acts as a Luttinger reservoir.  

\begin{figure}
\includegraphics[clip=true,width=9cm]{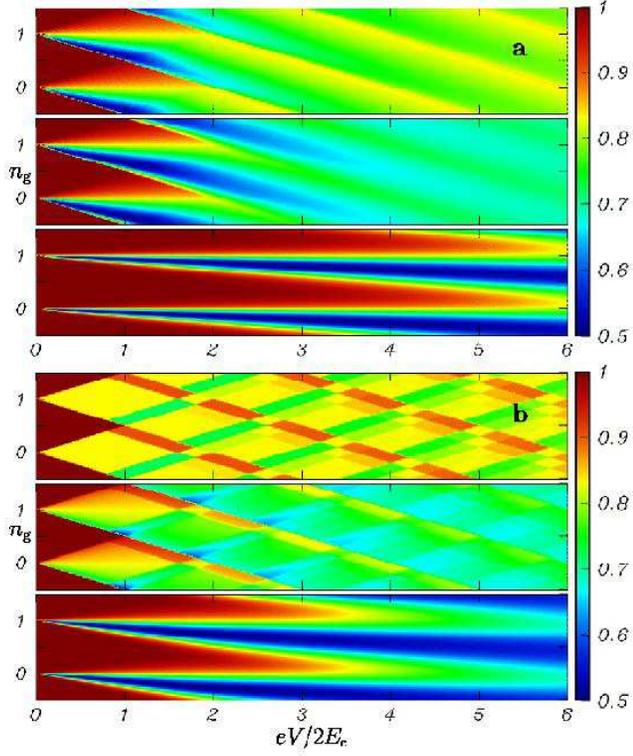}
\vskip-0.3cm
\caption{Density plots of the Fano factor $F=S/2e|\langle I\rangle|$
  in the $eV/2E_{\rm c}$, $n_{\rm g}$-plane for $R=10$, $T=0$ and 
  interaction strength $g=1,0.8,0.2$. The discretization energy are: 
  (a) $\varepsilon/E_{\rm
    c}=0.1$; (b) $\varepsilon/E_{\rm c}=1.6$. Color code (right).}
\label{braggiofig:1}
\end{figure}

In the following we will present the case of asymmetric barriers
characterized by ratio $R=R_{\rm r}/R_{\rm l}$. As we are interested
in the shot noise we consider only low temperatures $T\ll e V$. We
will consider results for the Fano factor $F\equiv S/2e|\langle
I\rangle|$. A detailed discussion of the behavior of the average
current as a function of the external voltages as been given in 
\cite{braggio01}.  The Fano factor is presented in Fig.
\ref{braggiofig:1} with $R=10$ and $T=0$ for two different ratio
$\varepsilon/E_{\rm c}=0.1$ (a) and $\varepsilon/E_{\rm c}=1.6$ (b).
These two values are chosen in order to show the different effects of
the discrete plasmon modes.  The noise is periodic in the $n_{\rm
  g}$-variable. For clarity in the figure we present the results for
$-1/2<n_{\rm g}<3/2$ where two Coulomb peaks of the linear
conductance, centered at $n_{\rm g}=0,1$, are present . Different
features of electron-electron interactions will be now described
depending on the small, large and intermediate voltage regions.  The
numerical results can be corroborated by analytical calculations in
several limits.

Let us start with the small voltage limit. Near to the resonance
$n_{\rm g}\approx 0$, for $eV<4E_{\rm c}$ two charge values
contribute predominantly to the transport. The Coulomb blockade
region are here delimitated by the diamond lines $E_{\pm}\equiv eV/2\pm
2E_{\rm c}n_{\rm g}=0$. In this regime the Fano factor assumes the form 
\begin{equation}
\label{fano0}
\!F=1\!-\frac{2R\gamma_{+}\gamma_{-}}
{[(1+e^{-E_{-}/T})\gamma_{-} + R(1+e^{-E_{+}/T})\gamma_{+}]^2}
\end{equation}
with $\gamma_{\pm}\equiv\gamma(E_{\pm})$.
By inspection of (\ref{fano0}) one can 
see that the temperature plays a significant role {\em only} near
the edges of the diamond lines $|E_{+}|\le T$ or $|E_{-}|\le T$ (see later). 
Out from this region the Fano factor is well approximated by the $T=0$ value.
We first discuss the case of low bias voltages $E_{\pm}<\varepsilon$ where 
it is not possible to excite plasmons. At $T=0$ the Fano is 
$F=(E_{-}^{\,2\nu(g)}+R^{2}E_{+}^{\,2\nu(g)})/(E_-^{\,\nu(g)}
+RE_{+}^{\,\nu(g)})^2$, with $\nu(g)=-1+1/g$.  For non-interacting
leads ($g=1$) $F$ is voltage independent $F=(1+R^2)/(1+R)^2$ and
assumes the lowest value ($F=1/2$) {\em only} for
symmetric barriers.

\begin{figure}
\includegraphics[clip=true,width=8cm]{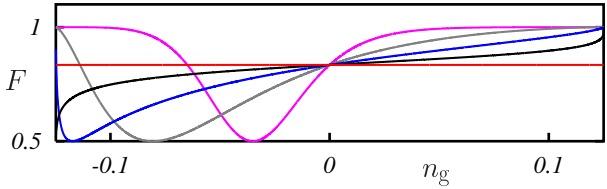}
\vskip-0.2cm
\caption{Fano factor  
as a function of the gate voltage $n_{\rm g}$,  for $T=0$, $R=10$,
$eV/2E_{\rm c}=0.25$, $\varepsilon/E_{\rm c}=1.6$, with interaction strength
$g=1$ red, $g=0.8$
  black, $g=0.6$ blue, $g=0.4$ grey, $g=0.2$ magenta.} 
\label{braggiofig:2}
\end{figure}

In the interacting case $g<1$ the Fano is {\em always} suppressed to
the $1/2$ value, at a given voltage ratio $4E_{\rm c}n_{\rm
  g}/eV=(1-R^{1/\nu(g)})/(R^{1/\nu(g)}+1)$ .  On increasing the
interaction strength ($g\to 0$) the resonant point moves towards
$n_{\rm g}=0$; away from it $F$ tends to the Poisson value (see
Fig.~\ref{braggiofig:1}(b) at $eV<4E_{\rm c}$,
and Fig.~\ref{braggiofig:2}).  This behaviour is also confirmed by the
temperature scaling near to the edges of the Coulomb blockade region.
In the limits $|E_{\pm}|\ll T$ one has
$$
F\approx 1-c_{\pm}(T/eV)^{\nu(g)}\,,
$$
with $c_{\pm} \sim R^{\pm 1}$.  For non-interacting leads, $F$ is
temperature independent, increasing the interactions the Fano factor 
approaches as a power law the values $F=1$. 
If $\varepsilon\ll E_{\rm c}$ (Fig.~\ref{braggiofig:1}(a)) one 
can also reach the regime $\varepsilon\ll E_{\pm}$, still using only 
two charge states, $eV<2E_{\rm c}$.
In this case the Fano factor results from the  contribution by the 
continuum of excited plasmons. Its expression as a function of resistance 
and voltages is still valid  but with the replacement 
$\nu(g)\to\nu(g/2)$. 

\begin{figure}
\vskip-0.5cm
\includegraphics[clip=true,width=8cm]{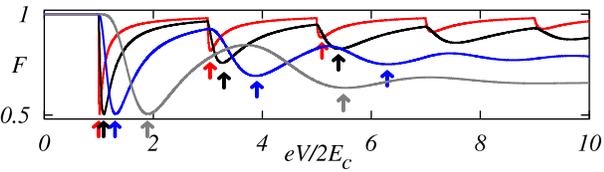}
\vskip-0.3cm
\caption{Fano factor as a function of the transport voltage 
$eV/2E_{\rm c}$, for $T=0$,
$R=100$, $n_{\rm g}=0.5$, $\varepsilon/E_{\rm c}=0.01$, with 
$g=1$ red, $g=0.8$
  black, $g=0.6$ blue, $g=0.4$ grey. The arrows 
 indicate the Fano dips whose distance at a given interaction 
(see different colors) follow equation~(\ref{deltaV0}).}
\label{braggiofig:3}
\end{figure}

The strong interplay between Coulomb blockade features and non-Fermi
liquid correlations is also very clearly revealed at larger voltages
$eV\gtrsim 4E_{\rm c}$, if the asymmetry of the barriers is not too
small.  As a function of the voltage bias and in the limit $eV \gg
\varepsilon$, (cf. Fig.~\ref{braggiofig:3}) the Fano factor shows a
series of dips and it saturates to a constant value, $F_{\infty}$,
for $e V\gg E_{\rm c}$.  In the orthodox theory of Coulomb blockade in
small metal junctions, the decrease of the Fano factor is driven by
$E_{\rm c}$ and it is attributed to the correlations present at the
degeneracy points.  A detailed analysis of the numerical results
allows to extract the following behaviour, valid for
enough strong asymmetry $R\gg 1$, for the distance in the bias voltage between
two successive dips at fixed voltage gate $\delta V_{\rm dip}$
\begin{equation}
\label{deltaV0}
\!\!\frac{e\delta V_{\rm dip}}{4E_{\rm c}}=
\frac{R^{1/\nu(g/2)}+1}{R^{1/\nu(g/2)}-1}\,,
\end{equation}
This fitting behaviour is
obtained by extrapolating, at larger voltages, the dip position lines
$eV(n)(1-R^{1/\nu(g/2)})=4E_{\rm c}(n_{\rm g}-n)(R^{1/\nu(g/2)}+1)$
founded previously, in the continuum case for the 
$n$ and $n+1$ charges, and considering the corresponding distance
$\delta V_{\rm dip}\equiv V(n+1)-V(n)$.  

In the asymptotic regime we find substancial modifications also to the
Fano-Schottky theorem $F_{\infty}=(1+R^2)/(1+R)^2$
~\cite{hershfield93}. We extract the power law behaviour
\begin{equation}
\label{deltaV}
F_{\infty}=\frac{1+R^{2/\nu(g/2)}}{[1+R^{1/\nu(g/2)}]^2}\,.
\end{equation}
We can see that Luttinger correlations lead, instead of $R$, a new
effective resistance ratio $R^{1/\nu(g/2)}$ that scales with a power
law in $g$, in agreement with the results at small voltages.
The properties of the Fano factor given so far, allow to extract the 
properties of the transistor and the value of the Luttinger liquid 
parameter without resorting to any fitting procedure.

Let us switch now to the more intricate region $eV\gtrsim 2E_{\rm c},
\varepsilon$, where Coulomb features, discrete nature of the excited
plasmon states, and Luttinger correlations are present simultaneously
(cf.  Fig.~\ref{braggiofig:1}(b)).  Apart from the asymptotic regime
$e V\gg E_{\rm c},\varepsilon$, where $F\to F_{\infty}$ as in 
(\ref{deltaV}), the physics is quite different depending on the
intensity of the interaction. For strong interactions, $g<1/2$, $F$ is
similar to the one described above in the continuum limit.
Qualitative differences are instead revealed in the weak interaction
limit $1/2<g<1$.  Here, the threshold of the Coulomb dips is {\em
  not} renormalized by the interaction. Moreover superimposed to these
dips there is an additional structure with a width proportional to
$|\varepsilon -2E_{\rm c}|$. These new features are due to the
transport via excited states.  Note that the effect of correlations in
these channels can also lead to a change of the curvature in the
Fano factor, differently from the processes associated with the ground
state to ground state transitions.  This is visible by looking at
the different colors of the diamont-like structures (e.g. red and
green features in Fig.~\ref{braggiofig:1}(b) for $g=1$).  These two
colors represent the two different way of entering/leaving in/from a
plasmon excited state of the dot.  Indeed, by considering the $n\to
n+1\to n$ transition, we can have either one electron that enters in
the dot, creating an excited plasmon state of the $n+1$ charge or one
elctron that leaves the dot from the $n+1$ ground state 
in an excited plasmon state of $n$ charges.  These two
processes can be discriminated in the Fano behaviour giving
informations on the different excited states that contribute to 
transport.

In conclusion we demonstrated how the presence of Luttinger
correlations changes drastically the
behaviour of the Fano factor with respect to the orthodox theory.  In
particular we showed that the interaction
strongly renormalizes the position of the dips, filtering at small
voltages the degrees of correlations.  A crucial role is played by the
barrier asymmetry that is shown to be power law renormalized by the
interaction strength. Very recently Nazarov and
Glazman~\cite{nazarov02} obtained the renormalization group equation
for the transmission in one-dimensional wire interrupted by a
double-barrier structure with a resonant level. It would be important
to extend the results on noise beyond the sequential tunnelling
approximation by emplying the approach of Ref.~\cite{nazarov02}.  We
expect that the predictions made in this work are amenable of
experimental verification. Indeed realizations of the one-dimensional
quantum dot system we considered has been recently achieved in a
cleaved-edge overgrowth quantum wire with two
impurities~\cite{auslaender00} and in carbon nanotubes with
buckles~\cite{postma01}.  Experimental results are already available
on the linear and non-linear transport current.  New experiments on
the shot noise would be of interest in order to test the new
predictions presented in this work.

Acknowledgements:
This work has been supported by Italian MURST via PRIN 2000, by the EU
within the programmes RTN2-2001-00440, HPRN-CT-2002-00144.

\begin{thebibliography}{99}

\bibitem{deJong97} 
        M. J. M. de Jong, and C. W. J. Beenakker, in {\em
        Mesoscopic Electron Transport}, edited by L. L. Sohn, L. P.
        Kouwenhoven, and G. Sch\"on (Kluwer Academic Publishers, Dordrecht,
        1997).

\bibitem{blanter00} 
        Y. M. Blanter, and M. B\"uttiker, Phys. Rep. {\bf 336}, 1 (2000).

\bibitem{depicciotto97}
        R. de Picciotto, M. Reznikov, M. Heiblum, V. Umansky, G. Bunin, 
        and D. Mahalu, Nature {\bf 389}, 162 (1997).

\bibitem{saminadayar97}
        L. Saminadayar, D. C. Glattli, Y. Jin, and B. Etienne,
        Phys. Rev. Lett {\bf 79}, 2526 (1997).

\bibitem{davies92} 
        J. H. Davies, P. Hyldgaard, S. Hershfield, and J. W. Wilkins, 
        Phys. Rev. B {\bf 46}, 9620 (1992).

\bibitem{hershfield93}
        S. Hershfield, J. H. Davies, P. Hyldgaard, C. Stanton, 
        and J. W. Wilkins, Phys. Rev. B {\bf 47}, 1967 (1993).

\bibitem{hanke93}
        U. Hanke, Yu. M. Galperin, K. A. Chao, and N. Zou,  
        Phys. Rev. B {\bf 48},  17209  (1993); 
        U. Hanke, Y. M. Galperin, and K. A. Chao, 
        ibid. {\bf 50},  1595  (1994).

\bibitem{korotkov94} 
        A. N. Korotkov, Phys. Rev. B {\bf 49}, 10381 (1994).

\bibitem{averin00}
        D. V. Averin, cond-mat/0010052.

\bibitem{sukhorukov00} 
        E. V. Sukhorukov, G. Burkard, and D. Loss, 
        Phys. Rev. B {\bf 63}, 125315 (2001).

\bibitem{birk95} 
        H. Birk, M. J. M. de Jong, and C. Sch\"onenberger, 
        Phys. Rev. Lett. {\bf 75}, 1610 (1995).

\bibitem{v95} 
        J. Voit, Rep. Progr. Phys. {\bf 58}, 977 (1995). 
  
\bibitem{kane92} 
        C. L. Kane, and M. P. A. Fisher, Phys. Rev. Lett. {\bf 68},
        1220 (1992); Phys. Rev. B {\bf 46}, 15233 (1992).

\bibitem{auslaender00} 
        O. M.  Auslaender, A. Yacoby, R. de Picciotto, K. W.
        Baldwin, L. N. Pfeiffer, and K. W. West, Phys. Rev. Lett. {\bf 84},
        1764 (2000).

\bibitem{bockrath99}
        M. Bockrath, D. H. Cobden, J. Lu, A. G. Rinzler, 
        R. E. Smalley, L. Balents, and P. L. McEuen, 
        Nature {\bf 397}, 598 (1999).

\bibitem{milliken96}
        F. P. Milliken, C. P. Umbach, and 
        R. A. Webb: Sol. State Comm. {\bf 97}, 309 (1996);
        A. M. Chang, L. N. Pfeiffer,  and K. W. West, 
        Phys. Rev. Lett. {\bf 77}, 2538 (1996).

\bibitem{chamon96}
        C. de C. Chamon, D. E. Freed, and X. G. Wen, 
        Phys. Rev. B {\bf 53}, 4033 (1996).

\bibitem{trauzettel02}
        B. Trauzettel, R. Egger, and H. Grabert, 
        Phys. Rev. Lett. {\bf 88}, 116401 (2002).


\bibitem{postma01} 
        H. W. Ch. Postma, T. Teepen, Z. Yao, M. Grifoni, and C. Dekker 
        Science {\bf 293}, 76 (2001).

\bibitem{braggio01} 
        A. Braggio, M. Sassetti, and B. Kramer, Phys. Rev. Lett.
        {\bf 87}, 146802 (2001).

\bibitem{k00} 
        T. Kleimann, M. Sassetti, B. Kramer, and A. Yacoby,
        Phys. Rev. B {\bf 62}, 8144 (2000).

\bibitem{nazarov02} Yu. V. Nazarov, and L. I. Glazman,
        cond-mat/0209090.
  \end{thebibliography}
\end{document}